\documentclass[twocolumn,aps,showpacs,prl] {revtex4}
\usepackage{amsmath}
\usepackage{graphicx}

\usepackage{dcolumn}
\usepackage{bm}
\usepackage{float}

\begin{document}

\title{Dynamic equivalence between atomic and colloidal liquids}
%\shorttitle{Dynamic equivalence between atomic and colloidal liquids} 
\author{Leticia L\'opez-Flores$^1$, Patricia Mendoza-M\'endez$^1$, Luis E.
S\'anchez-D\'iaz$^2$, Laura L. Yeomans-Reyna$^3$, Alejandro
Vizcarra-Rend\'on$^4$, Gabriel P\'erez-\'Angel$^5$, Mart\'in
Ch\'avez-P\'aez$^2$ and  Magdaleno Medina-Noyola$^2$}

%\shortauthor{Leticia L\'opez-Flores \etal}

\address{$^1$Facultad de Ciencias Fisico-Matem\'aticas,
Benem\'{e}rita Universidad Aut\'{o}noma de Puebla, Apartado Postal
1152, 72000 Puebla, Pue., M\'{e}xico}

\address{$^2$Instituto de F\'{\i}sica {\sl ``Manuel Sandoval Vallarta"},
Universidad Aut\'{o}noma de San Luis Potos\'{\i}, \'{A}lvaro
Obreg\'{o}n 64, 78000 San Luis Potos\'{\i}, SLP, M\'{e}xico}

\address{$^3$ Departamento de F\'{\i}sica, Universidad de Sonora, Boulevard
Luis Encinas y Rosales, 83000, Hermosillo, Sonora, M\'{e}xico.}

\address{$^4$Unidad Acad\'emica de F\'isica, Universidad Aut\'onoma de Zacatecas,
Paseo la Bufa y Calzada Solidaridad, 98600, Zacatecas, Zac., Mexico}

\address{$^5$Departamento de F\'isica Aplicada CINVESTAV-IPN, Unidad M\'erida
Apartado Postal 73 Cordemex  97310. M\'erida, Yuc., M\'{e}xico }

%{Structure of liquids. Time-dependent properties; relaxation}
%{Disperse systems; complex fluids. Colloids}

%\date{\today}

\begin{abstract}
We show that the kinetic-theoretical self-diffusion coefficient of
an atomic fluid plays the same role as the short-time self-diffusion
coefficient $D_S$ in a colloidal liquid, in the sense that the
dynamic properties of the former, at times much longer than the mean
free time, and properly scaled with $D_S$, will be indistinguishable
from those of a colloidal liquid with the same interaction
potential. One important consequence of such dynamic equivalence is
that the ratio $D_L/ D_S$ of the long-time to the short-time
self-diffusion coefficients must then be the same for both, an
atomic and a colloidal system characterized by the same
inter-particle interactions. This naturally extends to atomic fluids
a well-known dynamic criterion for freezing of colloidal liquids
[Phys. Rev. Lett. \textbf{70}, 1557 (1993)]. We corroborate these
predictions by comparing molecular and Brownian dynamics simulations
on the hard-sphere system and on other soft-sphere model systems,
representative of the ``hard-sphere" \emph{dynamic} universality
class.
\end{abstract}

\maketitle

One of the fundamental challenges in understanding the relationship
between dynamic arrest phenomena in colloidal systems
\cite{sciortinotartaglia}, and the glass transition in simple
glass-forming atomic liquids \cite{debenedetti}, is to determine the
role played by the underlying (Brownian vs. Newtonian) microscopic
dynamics. It is a widespread notion that colloidal systems
constitute a mesoscopic analog of atomic systems regarding the
relationship between inter-particle forces and macroscopic
properties \cite{pusey0,deschepperpusey1}. The molecular dynamics
simulation of an atomic liquid, for example, is expected to yield
the same equilibrium phase diagram, and a similar dynamic arrest
scenario, as the Brownian dynamics simulation of a colloidal liquid,
when referring to the same model system
\cite{lowenhansenroux,puertasaging}. Important questions, however,
remain unanswered, even at normal liquid states, far from the
neighborhood of the conditions for dynamic arrest. For example,
while it is well-known that monodisperse Brownian liquids will
freeze when the long-time self-diffusion coefficient $D_L$ reaches
about 0.1$\times D_S$, with $ D_S$ being the short-time
self-diffusion coefficient (``L\"owen's dynamic freezing criterion"
\cite{lowen}), no analogous criterion has been identified for the
corresponding atomic liquids.

In the attempt to develop the extension to \emph{atomic} liquids, of
the self-consistent generalized Langevin equation (SCGLE) theory of
\emph{colloid} dynamics \cite{todos2}, we have discovered that a
well-defined long-time dynamic equivalence between atomic and
colloidal liquids emerges upon the identification of the
kinetic-theoretical self-diffusion coefficient $D^0$ of an atomic
fluid \cite{chapmancowling}, as the analog of the short-time
self-diffusion coefficient $D_S$ of a colloidal liquid. In this
short communication we describe the physical foundations of this
extended SCGLE theory, which also constitute the physical basis of
the referred dynamic equivalence. One of the most important
manifestations of the latter is that the ratio $D^*\equiv D_L/ D_S$
must then be the same for an atomic and a colloidal system
characterized by the same inter-particle interactions, thus
naturally extending L\"owen's freezing criterion to atomic systems.
To corroborate these predictions we generate and compare molecular
and Brownian dynamics simulations on hard- and soft-sphere model
systems representative of the class of systems whose long-time
dynamics maps onto the dynamics of the hard-sphere fluid, i.e., that
pertain to the ``hard-sphere" \emph{dynamic} universality class
\cite{soft2}.

Let us start by considering a model atomic fluid, formed by $N$
spherical particles of mass $M$ in a volume $V$, interacting through
the pair potential  $u(r)$, whose microscopic dynamics is described
by Newton's equations. The fundamental concept upon which we
construct our theory for these properties is the role of the mean
free time, $\tau_0$, defined in the kinetic theory of gases as the
characteristic timescale that provides the crossover from the
short-time ballistic motion of the atoms to their long-time
diffusive transport. It is well known that for correlation times $t$
much shorter than $\tau_0$, and for distances much shorter than the
mean free path $l_{0}$,  all the particles move ballistically. For times $t$ much longer than $\tau_0$, each particle has undergone
many collisions, and its motion can be represented as a sequence of
random (ballistic) flights of mean length $l_{0}\sim 1/(n\sigma^2)$. As a consequence,
for $t\gg\tau_0$ each individual particle moves diffusively, with a
diffusion coefficient given by \cite{chandrasekhar} $D^0 = (l_0)^2/\tau_0= l_0v_0\sim \sqrt{k_BT/M}/(n\sigma^2)$ (where $l_0/\tau_0 =v_0\equiv (k_BT/M)^{\frac{1}{2}}$ is the thermal velocity). In fact, the rigorous value of
$D^0$, determined by the kinetic theory of gases
\cite{chapmancowling}, is
\begin{equation}
D^0\equiv \frac{3}{8\sqrt
\pi}\left(\frac{k_BT}{M}\right)^{1/2}\frac{1}{n\sigma^2}.
\label{dkinetictheory}
\end{equation}

This immediately implies that, on the average, the motion of each
individual particle crosses over from ballistic to diffusive, at the
crossover timescale $\tau_0$, and that this crossover behavior must
be described by the ordinary Langevin equation
\cite{mcquarrie,chandrasekhar} for the instantaneous velocity ${\bf
v}(t)$  of a representative tracer particle, $Md{\bf v}(t)/dt=
-\zeta^0 {\bf v}(t)+{\bf f}^0 (t)$. In this equation, the friction
coefficient $\zeta^0 $ is defined by Einstein's relation,
$\zeta^0\equiv k_BT/D^0$, with $D^0$ given by the
kinetic-theoretical result in Eq. (\ref{dkinetictheory}) above. This
strongly suggests a deeper analogy between the dynamics of Newtonian
and Brownian liquids, and hence, some form of dynamic equivalence
between atomic and colloidal dynamics, at least regarding tracer
diffusion phenomena. Thus, in both cases the relaxation time $\tau_0
= [M/\zeta^0]$ of the velocity, due to the friction force
$-\zeta^0{\bf v}(t)$, defines the crossover from ballistic
($t<<\tau_0$) to diffusive ($t>>\tau_0$) motion.

The fundamental difference lies, of course, in the physical origin
of the friction force  $-\zeta^0{\bf v}(t)$ and in the definition of
the friction coefficient  $\zeta^0$. In a colloidal liquid this
friction is caused by an ``external" material agent, namely, the
supporting solvent, which also acts as a heat reservoir. In an
atomic liquid, in contrast, the friction force $-\zeta^0{\bf v}(t)$
is not caused by any external agent, but by the spontaneous tendency
to establish, or restore, through molecular collisions, the
equipartition of the energy available for distribution  among  the
kinetic energy degrees of freedom of the system. Thus, in this case
the friction force $-\zeta^0{\bf v}(t)$ is the dissipative response
of the system towards the restoration of this partial thermal
equilibrium through collisional heat transport, and the spontaneous
fluctuations around this equipartition is the source of the
fluctuating force ${\bf f}^0 (t)$. As a consequence, the
corresponding value of $\zeta^0$ is determined by  the
kinetic-theoretical result for  $D^0$ in Eq. (\ref{dkinetictheory}),
through Einstein's relation $\zeta^0\equiv k_BT/D^0$.

The friction and fluctuating terms of the \emph{atomic} Langevin
equation  are,
of course, a consequence of the interatomic collisions of the tracer
particle with the rest of the particles in the fluid, an effect that
could also be described by the \emph{kinetic} terms of the pressure
tensor $\stackrel {\leftrightarrow}{\Pi} (\mathbf{r},t)$ of the
fluid formed by the surrounding particles. The
\emph{configurational} components of $\stackrel
{\leftrightarrow}{\Pi} (\mathbf{r},t)$, on the other hand,
correspond to the usual direct force
$-\nabla_T\sum_{i=1}^Nu(\mathbf{r}_{iT})$ (where ``T" stands for
``tracer"). This force, whose effects must still be taken into account, can also be written exactly as $\int d^3 r [\nabla u(\mathbf{r})] n(\mathbf{r},t)$, where $n(\mathbf{r},t)$ is the local density of particles at position $\mathbf{r}$ referred to the tracer particle's center \cite{faraday}. Thus, from the superposition of forces, the complete Langevin equation for a tracer particle in an atomic liquid actually reads $Md{\bf v}(t)/dt= -\zeta^0 {\bf v}(t)+{\bf
f}^0 (t)+\int d^3 r [\nabla u(\mathbf{r})] n(\mathbf{r},t)$ \cite{note1}. By writing the time-evolution equation of $n(\mathbf{r},t)$ (linearized around equilibrium), and by substituting the solution in this configurational contribution of the direct interactions, one can transform \cite{faraday}  this direct force term, into another friction term plus
its corresponding fluctuating force, namely,
\begin{equation}
M{\frac{d{\bf v}(t)}{dt}}= -\zeta^0{\bf v}(t)
+\textbf{f}^0(t)-\int_0^tdt'\Delta\zeta(t-t'){\bf v}(t') +
\textbf{F}(t). \label{gletracer}
\end{equation}
In this equation the term involving the time-dependent friction
function $\Delta\zeta(t)$ describes the mean dissipative friction
effects due to the collective direct force on the tracer particle,
whose random component is the Gaussian stationary stochastic force
$\textbf{F}(t)$. The derivation of this equation follows step by
step the derivation originally carried out for
colloidal liquids \cite{faraday}; it is now explained in detail for atomic fluids in Ref.  \cite{gleatomictracerdiff}. One of the main
products of such derivation is the following approximate but
general expression for $\Delta \zeta^* (t)\equiv\Delta
\zeta(t)/\zeta^0$, in terms of the static structure factor $S(k)$
and of the \emph{collective} and \emph{self} intermediate scattering
functions (ISF) $F(k,t)$, and $F_S(k,t)$ \cite{pusey0} of the fluid
surrounding the tracer particle,
\begin{equation}
\Delta \zeta^* (t) =\frac{D_0}{3\left( 2\pi \right) ^{3}n}\int d
{\bf k}\left[\frac{ k[S(k)-1]}{S(k)}\right] ^{2}F(k,t)F_{S}(k,t),
\label{dzdt}
\end{equation}
which is then also applicable in the present atomic case.

In order to evaluate $\Delta \zeta^*(t)$ we thus need to determine
$F(k, t)$ and $F_S(k,t)$ at least in the diffusive regime $t\gg
\tau^0$, since the effects described by $\Delta \zeta^*(t)$ in Eq.
(\ref{gletracer}) only manifest themselves at times longer than the
crossover time $\tau^0$. For this, one can employ the generalized
Langevin equation formalism \cite{faraday} to derive the
general memory function expressions for the ISFs of an atomic
liquid. Such an exercise is described in Ref. \cite{overdampedatomic}. As a result one derives general expressions for $F(k,t)$ and
$F_S(k,t)$, which in principle describe the dynamics of the atomic
liquid in the full time-domain, from the ballistic to the diffusive
regimes. Requesting consistency  of such general expressions, with
the physical picture leading to Eq. (\ref{gletracer}) above, these
expressions for $F(k,t)$ and $F_S(k,t)$ can be written in the
long-time (or ``overdamped") limit in terms of the corresponding
memory functions, $C(k,t)$ and $C_S(k,t)$. The latter may then be
approximated, as suggested in Ref. \cite{todos2}, by
$C(k,t)=C_S(k,t)=\lambda (k)\ \Delta \zeta^*(t)$, so that the
resulting long-time expressions for $F(k,t)$ and $F_S(k,t)$ can
finally be written, in Laplace space, as \cite{overdampedatomic}
\begin{equation}\label{fsdkz}
 F_S(k,z) = \frac{1}{z+\frac{k^2D^0 }{1+\lambda (k)\ \Delta \zeta^*(z)}}
\end{equation}
and
\begin{equation}\label{fdkz}
F(k,z) = \frac{S(k)}{z+\frac{k^2D^0 S^{-1}(k)}{1+\lambda (k)\ \Delta
\zeta^*(z)}},
\end{equation}
where the function $\lambda (k)$ is given by $\lambda (k)=1/[1+(
k/k_{c})^{2}]$, with $k_{c } = 1.35 (2\pi/\sigma)$.

For a given static structure factor $S(k)$, Eqs.
(\ref{dzdt})-(\ref{fdkz}) constitute a closed system of equations,
which turns out to be identical to the equations that summarize the
self-consistent generalized Langevin equation (SCGLE) theory of
\emph{colloid} dynamics \cite{todos2}. Thus, these equations, which
constitute an ``atomic" extension of the SCGLE theory, predict that
the long-time dynamic properties of an atomic liquid will then
coincide with the corresponding properties of a colloidal system
with the same $S(k)$, provided that the time is scaled as $D^0t$ and
that the respective meaning and definition of $D^0$ is taken into
account (i.e., that $D^0$ is given either by the short-time
self-diffusion coefficient in the case of the colloidal liquid, or
by Eq. (\ref{dkinetictheory}) in the case of the atomic fluid). One
obvious example of such long-time properties is the long-time
self-diffusion coefficient   $D_L \ \equiv \lim_{t \to \infty}
\langle(\Delta \textbf{r}(t))^2\rangle / 6t $, whose value   for an
atomic system, scaled as $D^*\equiv D_L/D^0 $, is thus predicted to
be indistinguishable from the corresponding property of the
equivalent colloidal system.

\begin{figure}
\begin{center}
\includegraphics[scale=.2]{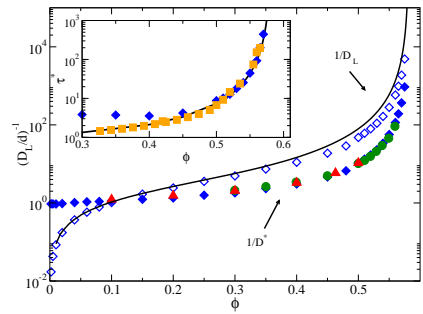}
\caption{Long-time self-diffusion coefficient $D_L (\phi)$ (main
figure), and dimensionless $\alpha$-relaxation time $\tau^*(k)\equiv
k^2D^0(\phi)\tau_{\alpha}(k;\phi)$ at $k\sigma=7.1$ (inset), of the
hard-sphere fluid. The results of the molecular
dynamics simulations described in Ref. \cite{gabriel} for $D_L
(\phi)$ are expressed in ``atomic units" (i.e., with $d=\sigma(k_BT/M)^{1/2}$, empty
diamonds), and scaled as  $D^* (\phi) \equiv
D_L(\phi)/ D^0(\phi)$, with
$D^0(\phi)$ given by Eq. (\ref{dkinetictheory}) (i.e., $d= D^0(\phi)$, full diamonds). The corresponding results for $\tau^*(k)$ are the full diamonds in the inset. The other full symbols
are the \emph{Brownian} dynamics simulation results for $D^*$ from Refs.
\cite{cichocki} (triangles) and \cite{tokuyamabd} (circles), and for
$\tau^*$ from our present BD simulation results for HS using the
method described in Ref. \cite{dynamicequivalence} (squares). The
solid lines correspond to the solution of the SCGLE theory.}
\label{fig1}
\end{center}
\end{figure}

This is an important and distinct prediction of the SCGLE theory,
whose accuracy can be readily checked without even solving the
corresponding self-consistent approximate equations above, since one
can compare molecular and Brownian dynamics results for the
long-time self-diffusion coefficient of a given specific system.
Thus, in Fig. \ref{fig1} we plot molecular dynamics data for
$D_L(\phi)$ of a hard-sphere fluid both, in the ``usual" atomic
units $ \sigma(k_BT/M)^{1/2}$, and scaled as $D^* (\phi) \equiv
D_L(\phi)/ D^0(\phi)$, with $D^0(\phi)$ given by Eq.
(\ref{dkinetictheory}). The same figure also presents available
Brownian dynamics simulation results for $D_L(\phi)$ of the hard
sphere system without hydrodynamic interactions, scaled as $D^*
(\phi)\equiv D_L(\phi)/ D^0$, with $D^0$ being the
$\phi$-independent short-time self-diffusion coefficient of the
Brownian particles. Clearly, the ``colloidal" and the ``atomic"
results for $D^*$ collapse onto the same curve, which we denote by
$D^*_{HS} (\phi)$. One immediate and important consequence of this
comparison is the extension to atomic liquids of L\"owen's dynamic
criterion for freezing: now for both, the atomic and the colloidal
HS fluids, the condition $D^*_{HS} (\phi)\approx 0.1$ occurs at
$\phi=\phi_{HS}^{(f)}=0.494$.

This long-time colloidal--atomic correspondence is also observed in
the $\alpha$-relaxation time $\tau_{\alpha}(k;\phi)$, defined by the
condition $F_S(k,\tau_{\alpha})=1/e$, but only in the metastable
liquid regime. This is illustrated in the inset of Fig. \ref{fig1},
which demonstrates that the molecular dynamics and the Brownian
dynamics data for $\tau^*(k)\equiv k^2D^0\tau_{\alpha}(k;\phi)$
evaluated at $k\sigma=7.1$ collapse onto a common curve in the
metastable fluid regime, $0.5 \lesssim \phi$, but for $\phi$ below
freezing, both data depart from each other. The reason for this is
that in reality, for $\phi \lesssim 0.5 $, the dynamics of
$F_S(k,t)$ at times $t\approx \tau_{\alpha}$ is not yet described by
the long-time asymptotic expressions in Eqs. (\ref{fsdkz}) and
(\ref{fdkz}); instead, $F_S(k,t)$ is closer to its ballistic
approximation $F_S(k,t)\approx \exp[-k^2v_0^2t^2/2]$, so that
$\tau_{\alpha}(k;\phi)\approx \sqrt{2}/kv_0$ (for colloidal
fluids, $\tau_{\alpha}(k;\phi)\approx 1/k^2D^0$ at low densities).

Let us also point out that besides comparing MD vs. BD simulation
data, one can, of course, also solve Eqs.(\ref{dzdt}),
(\ref{fsdkz}), and (\ref{fdkz}) to actually calculate theoretically
$D^*_{HS} (\phi)$ and $\tau^*(k)$. The solid lines in Fig.
\ref{fig1} are the result of such numerical calculation, with $S(k)$
provided by the Percus-Yevick approximation \cite{percusyevick} with
its Verlet-Weis correction \cite{verletweis}. In fact, let us
mention that the data for $\tau^*(k)$ in the inset were employed to
calibrate the parameter $k_c$ in the function $\lambda (k)$, thus
leading to the quoted value $k_c=1.35(2\pi/\sigma)$.

\begin{figure}
\begin{center}
\includegraphics[scale=.2]{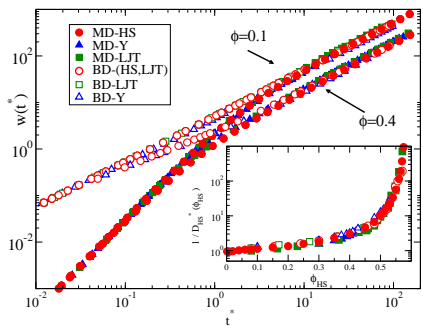}
\caption{Main figure: Scaled mean squared displacement $w(t^*)
\equiv \ <(\Delta \textbf{r}(t))^2>/6l_0^2 $ as a function of the
dimensionless time $t^*\equiv t/\tau_0$, simulated with molecular
dynamics (MD, solid symbols) and Brownian dynamics (BD, empty symbols) for
fluids with hard-sphere (HS, circles), truncated Lennard-Jones (TLJ, squares)
and repulsive Yukawa (Y, triangles) interactions, for states
corresponding to $\phi_{HS}$ = 0.1 and 0.4. Inset: simulation data for the scaled long-time self-diffusion coefficient
$D^*$ of the same systems (and same symbols), plotted as a function
of the effective HS volume fraction $\phi_{HS}=\phi_{HS} (n,T)$.
 } \label{fig2}
\end{center}
\end{figure}

Another manner to express this long-time dynamic universality of
atomic and colloidal liquids is to plot the simulated data of the
dimensionless MSD $w(t^*) \equiv \ <(\Delta \textbf{r}(t))^2>/6l_0^2
$ as a function of the dimensionless time $t^*\equiv t/\tau_0$. From
the generalized Langevin equation (GLE) in Eq. (\ref{gletracer}) one
can derive the following equation for $w(t^*)$
\begin{equation}
\frac{dw(t^*)}{dt^*} +w(t^*) =t^*-\tau^0\int_0^{t^*} \Delta \zeta^*
(t^*-t^{*'})w(t^{*'})dt^{*'}, \label{wdtintdifec}
\end{equation}
whose solution satisfies the short- and long-time limits  $w(t^*\to
0) \approx t^{*2}/2$ and  $w(t^*\to \infty) \approx D^* t^*$, with
$D^*=1/[1+\Delta \zeta^*(z=0)]$. This scaling, however, hides the true timescales
for atomic and colloidal systems, which differ by many orders of magnitude \cite{gleatomictracerdiff}.
For example, the cross-over time $\tau_0$ from
ballistic to diffusive motion is of the order of one picosecond in an atomic liquid and of tens of nanoseconds in a typical colloidal system, whereas the crossover time from short to long time diffusion,  $\tau_I\equiv d^2/D^0$ (with $d=n^{-1/3}$), is of the order of tens of picoseconds in an atomic liquid and of a fraction of a second in a colloidal liquid. Thus, in a colloidal fluid $\tau_I \approx 10^7 \tau_0$, and hence, the ballistic short-time regime is generally unimportant. Thus, one normally
redefines ``short times" by first taking the ``overdamped" limit
(i.e., dropping the inertial term in Eqs. (\ref{gletracer}) and
(\ref{wdtintdifec})). This only changes the short-time limit to
$w(t^*) \approx t^*$, but leaves unaltered the long-time limit
$w(t^*) \approx D^* t^*$. Thus, except for these short-time
differences, the msd of an atomic and a colloidal liquid with the
same interactions and the same  $S(k)$ should be indistinguishable
at long times. This is precisely what is illustrated in Fig.
\ref{fig2}, which plots the simulated data of $w(t^*)$  as a
function of  $t^*$ for the hard-sphere system at two volume
fractions, $\phi$ = 0.1 and 0.4.

Fig. \ref{fig2}  also illustrates the fact that this
colloidal--atomic dynamic correspondence is not restricted to the
hard-sphere fluid, but it actually extends over to systems with soft
repulsive interactions. This is a direct result of combining the
present colloidal--atomic correspondence for the hard sphere system,
with another important scaling rule, which derives from the
principle of dynamic equivalence between soft-sphere and hard-sphere
liquids \cite{soft2}. This principle states that the dynamic
properties of a colloidal liquid at number concentration $n$ and
temperature $T$, whose particles interact through a repulsive
soft-sphere pair potential $u(r)$, will be given by those of the
hard-sphere system with an effective hard-sphere volume fraction
$\phi_{HS}(n,T)$ determined by the iso-structurality condition,
$S(k_{mak};n,T)=S_{HS}(k^{(HS)}_{mak};\phi_{HS})$, which requests
that the height of the main peak of the static structure factor of
the ``real" soft-sphere system, and of the effective hard-sphere
system, coincide. As a consequence one has, for example, that the
msd scales as $w(t^*;n,T) = w_{HS}(t^*;\phi_{HS}(n,T))$ and that the
curve $D^*_{HS} (\phi)$ in Fig. \ref{fig1} becomes a universal curve
for all colloidal soft-sphere liquids, provided that its horizontal
axis refers to the effective volume fraction $\phi_{HS} (n,T)$.

The extension of this scaling to atomic systems is immediate once
the collision diameter $\sigma$ entering in the expression for
$D^0(n,T)$ in Eq. (\ref{dkinetictheory}) is given a proper
definition for the soft-sphere potential $u(r)$ considered. We found
that replacing this collision diameter by $\sigma_{HS}(n,T)\equiv
[6\phi_{HS}(n,T)/\pi n]^{1/3}$ seems to be a simple and reasonably
universal approximation. To assess the accuracy of the resulting
general scaling, we generated molecular and Brownian dynamics
simulation data for the truncated Lennard-Jones system, $
u(r)/k_BT=T^{*-1}\left[(\sigma/r )^{12}
-2(\sigma/r)^6+1\right]\theta (\sigma-r)$, at fixed reduced
temperature $T^*=1$, and for the repulsive Yukawa potential $
u(r)/k_BT= K \exp [-z(r/\sigma-1)]/(r/\sigma)$, with $K=554$ and
$z=0.149$. In both cases, the densities were chosen to correspond to
effective HS systems at volume fractions $\phi_{HS}(n,T)$ = 0.1 and
0.4. The results for $w(t^*;n,T)$ in Fig. \ref{fig2} illustrate
that, plotted in this scaled manner, the long-time limit of
$w(t^*;n,T)$ will not discriminate between atomic or colloidal
systems and between soft- and hard-sphere interactions. Thus, from
any of these simulations one should be able to determine $D^*$. To
illustrate this, in the inset of Fig. \ref{fig2} we plot the
simulated data of $D^*(n,T)$ for these two soft-sphere model systems
as a function of $\phi_{HS} (n,T)$, which clearly collapse onto the
curve $D^*_{HS} (\phi_{HS})$, represented in the figure by the HS MD
data.

In summary, we have shown that at least for model liquids whose
structure is dominated by (soft- or hard-sphere) repulsive
interactions, the long-time dynamics of atomic liquids is
indistinguishable from the dynamics of the colloidal systems with
the same inter-particle interactions. As a consequence, just like
the equilibrium thermodynamic and structural properties, some
dimensionless long-time dynamic properties, such as $D^*$ and
$\tau^*(k)$ (the latter only in the supercooled liquid regime), will
exhibit the same independence from the short-time microscopic
dynamics which otherwise distinguishes atomic from colloidal
systems. It is reasonable to expect that this dynamic universality
will be useful in understanding, for example, the relationship
between dynamic arrest phenomena in colloidal systems, and the glass
transition in simple glass-forming atomic liquids.

\bigskip

\acknowledgments

This work was supported by the Consejo Nacional de
Ciencia y Tecnolog\'{\i}a (CONACYT, M\'{e}xico) (grants 84076 and
132540) and by Fondo Mixto CONACyT-SLP (grant
FMSLP-2008-C02-107543).


\begin{thebibliography}{0}
\bibitem{note1} Strictly speaking, this expression is only valid for continuous potentials, the hard-sphere system being understood as the infinitely rigid limit of a soft-sphere continuous potential with tunable softness.
\end{thebibliography}
\end{document}